\documentclass[twocolumn,         
               showpacs, showkeys,           
               preprintnumbers,     
               aps,                 
               prd,          	    
               letterpaper,             
               nofootinbib,         
               tightenlines,        
               floats,floatfix      
               ]{revtex4-1}
\usepackage{dcolumn}   
\usepackage{bm}        
\usepackage{float}

\usepackage{graphicx,color}
\usepackage{latexsym}
\usepackage{cancel}
\usepackage{amsmath,amssymb}        
\usepackage[colorlinks=true,linkcolor=blue,citecolor=blue]{hyperref}
\usepackage{mathrsfs}
\usepackage{comment}
\usepackage{soul}
\definecolor{purple}{rgb}{0.58,0.0,0.83}
\definecolor{orange}{rgb}{1,0.5,0}
\DeclareSymbolFontAlphabet{\mathrsfs}{rsfs}
\DeclareMathAlphabet{\mathcal}{OMS}{cmsy}{m}{n}

\newcommand{\scri}{\mathrsfs{I}}

\usepackage{verbatim} 

\begin{document}


\title{Frequency shift of light emitted from growing and shrinking black holes}


\author{F. G. Guzm\'an}
\affiliation{Instituto de F\'{\i}sica y Matem\'{a}ticas, Universidad
              Michoacana de San Nicol\'as de Hidalgo. Edificio C-3, Cd.
              Universitaria, 58040 Morelia, Michoac\'{a}n,
              M\'{e}xico.}
\author{I. Alvarez-R\'ios}
\affiliation{Instituto de F\'{\i}sica y Matem\'{a}ticas, Universidad
              Michoacana de San Nicol\'as de Hidalgo. Edificio C-3, Cd.
              Universitaria, 58040 Morelia, Michoac\'{a}n,
              M\'{e}xico.}
\author{J. A. Gonz\'alez}
\affiliation{Instituto de F\'{\i}sica y Matem\'{a}ticas, Universidad
              Michoacana de San Nicol\'as de Hidalgo. Edificio C-3, Cd.
              Universitaria, 58040 Morelia, Michoac\'{a}n,
              M\'{e}xico.}


\date{\today}


\begin{abstract}
In this paper we present a method to study the frequency shift of signals sent from near a Schwarzschild black hole that grows or shrinks through accretion. We construct the numerical solution of Einstein's equations sourced by a spherical shell of scalar field, with positive energy density to simulate the growth and with negative energy density to simulate the shrink of the black hole horizon. We launch a distribution of null rays at various time slices during the accretion and estimate their energy along their own trajectories. Spatially the bundles of photons are distributed according to the distribution of dust, whose dynamics obeys Euler equations in the test field limit during the evolution of the black hole. With these elements, we construct the frequency shift of photons during the accretion process of growth or contraction of the hole, which shows a variability that depends on the thickness of the scalar field shell or equivalently the time scale of the accretion.
\end{abstract}


\keywords{redshift and velocities -- numerical simulations in gravitation and astrophysics}


\maketitle

\section{Introduction}

The recent results on direct observations of black holes have boosted the intensity of studies related to the observable astrophysical processes in the black hole's near horizon region \cite{2019ApJ...875L...1E}. Observational efforts are also pushing the boundaries of resolution and eventually  future  observations could be able to track events within time scales of black hole mass, for example variability time-scales (e. g. \cite{2018ApJ...864....7M}).

In this paper we present a method to calculate signatures of growing and shrinking of black holes. For this we consider a spherically symmetric black hole that can grow or shrink, and study the frequency shift of light emitted from near the black hole's horizon and how it would be  observed from far away. The growing and shrinking processes are simulated by solving Einstein's equations coupled to a scalar field.  We use a {\it regular scalar field} with positive energy density to produce growth (see e.g. \cite{Thornburg:1999iy,GuzmanLora2013}) and a {\it phantom scalar field} that violates the weak energy condition in order to produce a decrease of the black hole mass (see e.g.\cite{GonzalezGuzman2009,GonzalezGuzman2016}). 

In order to study the effects of growing/shrinking event horizon on test matter around the black hole, which is eventually the source of light to be observed, we also solve Euler's equations for a dust fluid in the test field approximation during the evolution of the space-time.  At various time slices during the process of growth/shrink of the black hole horizon, we launch a bundle of photons distributed according to the distribution of the dust, which is also time-dependent. We  calculate the energy of photons along their null rays themselves and estimate the energy shift along these trajectories.

In our analysis we restrict the parameter space to the accretion of scalar field shells, whose mass energy is one quarter of the initial black hole horizon mass and thickness of size from one to ten times the horizon mass. The thiner the shell is, the faster the accretion, and in our parameter space the accretion time scale from 20 to 100 units of time in terms of the black hole mass. With these parameters the photon frequency can be red-shifted up to 60\% during the process of growth of the hole, or blue-shifted during a shrinking process. These values are specific for our parameter space, but the method is generic for an arbitrary accretion time scale.

Among the astrophysical motivations to carry on the analysis in this paper and potential applications of the method, we have the case of ultralight bosonic dark matter, where our method could be useful to set predictions or limitations to this dark matter model through the effects on accretion onto supermassive black holes (e.g. \cite{PhysRevD.102.063022,chung2021searching,deluca2021tidal,marsh2021astrophysical}). Another possibility is to extend the analysis and estimate the variability in slow processes like Hawking radiation (see e.g. \cite{Marto_2021}), or estimate predictions on the accretion of exotic dark energy that may eventually decrease a black hole size as well (e.g. \cite{2013Galax...1..114F,2018Univ....4..109F,2015IJMPD..2450061J}).

The paper is organized as follows. In Section \ref{sec:equations} we present the model and the equations governing the evolution of the black hole, the scalar field and the perfect fluid. In Section \ref{sec:redshift} we describe how we analyze the red/blue shift of light sent from near the black hole. In Section \ref{sec:results} we present the results on frequency shift of photons. Finally in Section \ref{sec:comments} we draw some final comments.

\section{Equations of space-time and matter}
\label{sec:equations}

\subsection{Einstein's equations}

We start by assuming the black hole space-time is governed by the Lagrangian density

\begin{equation}
{\cal L} = -R + \frac{\kappa}{2} g_{\mu\nu} \partial^{\mu}\phi \partial^{\nu}
\phi + V(\phi) + L_{pf},\label{eq:Lagrangian}
\end{equation}

\noindent where $R$ is the Ricci scalar of the space-time, $g_{\mu\nu}$ is the space-time metric, $\phi$ is the scalar field and $V(\phi)$ is the scalar field potential, that for our analysis we set to zero, which in turn means that we use  massless scalar fields. The term $L_{pf}$ is the Lagrangian density of a perfect fluid. 
The resulting system of equations is the Einstein-Klein-Gordon-Perfect fluid system, consisting in Einstein's equations, Klein-Gordon equation, mass conservation of the perfect fluid and the local conservation of the stress-energy tensor of the perfect fluid that read:

\begin{eqnarray}
&&G_{\mu\nu} = T^{\phi}_{\mu\nu}+\cancel{ T^{pf}_{\mu\nu}   },\label{eq:EKG}\\
&&\frac{1}{\sqrt{-g}}\partial_{\mu}
        [\sqrt{-g}g^{\mu\nu} \partial_{\nu}\phi] = -\partial_{\phi}V,\label{eq:KG}\\
&& \nabla_{\mu}(\rho_0 u^{\mu}) = 0 \label{eq:GRelEulerConsMass},\\
&&\nabla^{\mu}T^{pf}_{\mu\nu} =  0 \label{eq:GRelEulerCovarianteqs},
\end{eqnarray}

\noindent where $G_{\mu\nu}$ is Einstein tensor, matter is described by the stress-energy tensor for the scalar field

\begin{equation}
T^{\phi}_{\mu\nu} = \kappa \partial_{\mu} \phi \partial_{\nu}\phi -\frac{1}{2}g_{\mu\nu}[\kappa\partial^{\alpha}\phi \partial_{\alpha}\phi + 2V],\label{eq:setSF}
\end{equation}

\noindent  where $\kappa=\pm 1$ is the parameter that distinguishes between a {\it regular scalar field} with $\kappa=1$ and a {\it phantom} scalar field with $\kappa=-1$; in the later case the stress-energy tensor violates the null energy condition $T_{\mu\nu}k^{\mu}k^{\nu} \le 0$, where $k^{\mu}$ is a null vector, which implies the violation also of the weak energy condition and observers following time-like trajectories can measure negative energy densities. In this case there are unusual implications in astrophysical scenarios, because the area increasing theorem does not apply in this case \cite{hawking1975large}.The perfect fluid is represented by the tensor

\begin{equation}
T^{pf}_{\mu\nu}=\rho_0 h u_{\mu}u_{\nu}+ pg_{\mu\nu}\label{eq:setPF}
\end{equation}

\noindent where each fluid volume element has  rest mass density $\rho_0$, specific enthalpy $h=1+e+p/\rho_0$, internal energy $e$, pressure $p$ and  4-velocity $u^{\mu}$. We want the fluid to be a test field, which is achieved with the cancellation of the second term in Einstein's equations (\ref{eq:EKG}) whereas (\ref{eq:GRelEulerConsMass}) and (\ref{eq:GRelEulerCovarianteqs}) are solved for the space-time determined by the Einstein-Klein-Gordon (EKG) subsystem.

\subsection{Numerical solution}
\label{subsec:gbssn}

For the numerical solution of the EKG subsystem we assume the space-time to be spherically symmetric,  and therefore the accretion flux is assumed to be spherical as well.
The numerical relativity approach to this problem is well known and offers various options. Here we follow the approach in \cite{GonzalezGuzman2009,GonzalezGuzman2016} that we summarize here, and that allows the simulation of the black hole growing and shrinking scenarios. We assume the 3+1 decomposition of space-time, whose metric we write as

\begin{eqnarray}
ds^2 &=& -\left( \alpha^2 - \beta^r \beta^r \frac{g_{rr}}{\chi} \right)dt^2
         + 2\beta^r \frac{g_{rr}}{\chi}dtdr  \nonumber\\
        &+& \frac{1}{\chi} \left[g_{rr} dr^2 + g_{\theta\theta}
        (d\theta^2 + \sin^2 \theta d\varphi^2)\right], \label{eq:metric}
\end{eqnarray}

\noindent where $\beta^r$ is the only nonzero component of the shift vector, $\alpha$ is the lapse function
and $\chi$ acts as a conformal factor relating this metric to a spatial flat metric. Einstein's equations are solved as an   Initial Value Problem with the black hole constructed using puncture type of initial data \cite{PhysRevLett.78.3606}. The evolution uses the  Generalized Baumgarte Shapiro Shibata Nakamura (GBSSN) formulation \cite{Brown}, which for metric (\ref{eq:metric}) reduces to evolution equations for $g_{rr}$,  $g_{\theta\theta}$, the nonzero trace-free part
of the conformal extrinsic curvature $A_{rr}$, the trace of the
extrinsic curvature $K$ and the contracted conformal Christoffel nonzero symbol
$\Gamma^{r}$ as shown in \cite{Brown,PhysRevD.59.024007}.

The gauge evolves according to the 1+$\log$ slicing condition $\partial_t \alpha = \beta^a \partial_a
\alpha - 2\alpha K$ and the $\Gamma$-driver $\partial_t \beta^a = \frac{3}{4}B^a + \beta^c \partial_c \beta^a$, where 
$\partial_t B^a = \partial_t \Gamma^a + \beta^c \partial_c B^a - \beta^c \partial_c \Gamma^a - \eta B^a$, useful conditions for the evolution of black holes that avoid slice stretching effects. Unlike in \cite{GonzalezGuzman2009,GonzalezGuzman2016}, we use excision within the apparent horizon of the black hole, since we found that excision-without excision triggers  instabilities on the fluid near the horizon.

{\it Scalar field.} For the solution of the Klein-Gordon equation we write the equation (\ref{eq:KG}) as a first-order system in time for the auxiliary variables $\psi := \partial_r \phi$ and $\pi:=\sqrt{\gamma}\left(\partial_t\phi - \beta^r \partial_r\phi\right)/\alpha$, where $\gamma$ is the determinant of the metric of space-like hypersurfaces of the 3+1 metric (\ref{eq:metric}).

For the evolution of geometry and scalar field we use the method of lines with a fourth-order Runge-Kutta time integrator, with  fourth order accurate spatial discretization, and a permanent diagnostics of the GBSSN constraints \cite{Brown}. The initial conditions used fulfill these constraints at initial time, after a scalar field profile is prescribed, and then we produce a free evolution.

{\it Perfect fluid.} For the solution of the continuity equation (\ref{eq:GRelEulerConsMass}) and Euler equations of the fluid (\ref{eq:GRelEulerCovarianteqs}) we write down this subsystem in a flux balance law form, with conservative variables $D=\sqrt{\gamma}\rho_0 W$, $J_r=\sqrt{\gamma}\rho_0 h W^2 v_r$ and $\tau=\sqrt{\gamma}(\rho_0 h W^2  -p - \rho_0 W)$, where $v^r=\frac{u^r}{u^0}$ is the 3-velocity radial component of a volume element and $W=1/\sqrt{1-v^r v_r}$ its Lorentz factor. In general it is easy to assume the fluid obeys an ideal gas equation of state, however we consider only the simple case of a pressure-less gas, and therefore in what follows we set $p=0$.

For the solution we use a finite volume discretization and  high resolution shock capturing methods with linear  reconstructors and the  Harten, Lax, van Leer, Einfeldt approximate flux formula (HLLE) \cite{hlle,hlle2}  as illustrated in \cite{rezzolla2013relativistic,RMFEducativo}. The time update is done using the method of lines at the same time as for geometry and the scalar field.

The equations for geometry, scalar field and fluid, are solved on a finite domain  ${\cal D}:=r\in [r_{min},r_{max}]\times t\in[0,t_f]$, where $r_{min}=r_{AH}^0/2$ is the excision boundary and $r_{max}$ is the external boundary located in all cases at $r_{max}=200r_{AH}^0$ where $r_{AH}^0$ is the radius of the apparent horizon of the black hole at initial time. For the numerical solution we define a numerical domain ${\cal D}_d = \left\{ (r_i,t^n) \in {\cal D}~ |~ r_i=r_{min}+i\Delta r, t=n\Delta t \right\}$, where $\Delta r=(r_{max}-r_{min})/N_r$ and $\Delta t =C \Delta r$ are the spatial and time resolutions,  with $C$ the Courant-Friedrichs-Levy factor, and $N_r$ is the number of cells that define teh spatial discretization of ${\cal D}_d$. The base resolution used is $\Delta r = M^0_{AH}/25$ and $CFL=0.25$ with $M^0_{AH}$ the mass of the apparent horizon at  initial time.

\subsection{Initial conditions}
\label{subsec:id}

\textit{Scalar field.} At initial time we provide the scalar field profile through the variables $\phi(r,t=0)=\phi_0(r)$, $\psi(r,t=0)=\psi_0(r)$ and $\pi(r,t=0)=\pi_0(r)$. We choose the following scalar field profile 

\begin{equation}
\phi_0(r) = \dfrac{f(r)}{r} = \frac{A}{r}e^{-\dfrac{(r-r_0)^2}{\sigma^2}},
\label{eq:IC_phi}
\end{equation}

\noindent which is a Gaussian shell pulse launched toward the black hole. From this profile $\psi_0(r)$ is simply the derivative with respect to $r$ of $\phi_0(r) $, and finally we set  $\partial_t \phi |_{t=0} = \frac{1}{r} \frac{df}{dr}$, which corresponds to the inward pulse of the general solution for a spherical wave,  which in turn fixes the initial conditions for $\pi_0(r)$. Once this profile is prescribed it is necessary to solve the GBSSN constraints at initial time as follows. 

\textit{Solution of constraints for the space-time quantities.} We describe the space-time of the Schwarzschild black hole using modified isotropic coordinates. The ansatz we use for the metric components reads

\begin{equation}
g_{rr}=1,\qquad g_{\theta\theta}=r^2,\qquad \chi = \left(1+\dfrac{M}{2r}+u\right)^{-4},
\label{eq:metric_id}
\end{equation}

\noindent where $M$ is the ADM mass of the space-time, whereas $u=u(r,t=0)$ is a function to be determined through the solution of the Hamiltonian constraint at initial time, which reads

\begin{equation}
 u^{\prime\prime} = \dfrac{4\rho^{\phi}+3A_{rr}^2}{16 \chi^{5/4}}-2\dfrac{u'}{r^2},
\label{eq:u_}
\end{equation}

\noindent with $\rho^{\phi} = n^\mu n^\nu T^{\phi}_{\mu\nu}$ is the scalar field density. For the extrinsic curvature we set

\begin{equation}
K=0, \qquad A_{rr} = A_{rr}(r,t=0),
\end{equation}

\noindent where the $A_{rr}$ component is determined through the momentum constraint written as

\begin{equation}
A^\prime_{rr} + \dfrac{3\left[r+(M-2r^2 u^\prime )\chi^{1/4}\right]}{r^2}A_{rr}=S_r,
\label{eq:Arr}
\end{equation}

\noindent  with $S_r= \gamma_r{}^\mu n^\nu T_{\mu\nu}^\phi$ is the mixed projection of the stress-energy tensor of the scalar field. We solve the system (\ref{eq:u_}), (\ref{eq:Arr}) using a fourth order accurate ordinary differential equation integrator with the initial conditions $u(r_{min})=u^\prime(r_{min})=A_{rr}(r_{min})=0$. 

Finally, the gauge at initial time uses a pre-collapsed lapse of the form $\alpha=(1+M/2r)^{-1/2}$, zero shift
$\beta^r=0$ and $B^r=0$.

\textit{Perfect Fluid.} In order to study a test fluid we consider a pressure-less perfect fluid, whose stationary distribution in the coordinates of (\ref{eq:metric}) reads:

\begin{eqnarray}
\rho_0 &=& - \dfrac{C_1 \chi}{r^2 \sqrt{1-\alpha^{2}}},
\label{eq:IC_rho}\\
v^r &=& -\alpha\sqrt{\left(\alpha^{-2}-1\right)\chi}~,
\label{eq:IC_v}
\end{eqnarray}

\noindent for the rest mass density and radial velocity field, where $C_1$ is an integration constant. These functions can be made explicit once the conformal factor $\chi$ has been found by solving the constraints (\ref{eq:u_}) and (\ref{eq:Arr}) at initial time.

\subsection{Numerical methods and diagnostics}

{\it Apparent horizon location}. We track the location of the apparent horizon as the outermost  marginally trapped surface. For the metric (\ref{eq:metric}) it is equivalent to locate the outermost zero of the equation of the expansion of null spheres given by

\begin{eqnarray}
\Theta &=& \frac{1}{g_{\theta\theta}\sqrt{\chi g_{rr}}}
        \left(\chi (\partial_r g_{\theta\theta}
        )  - g_{\theta\theta}  (\partial_r \chi)\right) \nonumber\\
        &+& 2 \left( \frac{A_{rr}}{2g_{rr}} - \frac{1}{3} K \right),
\label{eq:ah}
\end{eqnarray}

\noindent at every time step, found at the coordinate radius $r=r_{AH}$. We  calculate the mass of the apparent 
horizon as $M_{AH} = R_{AH}/2$, where $R_{AH}=\sqrt{g_{\theta\theta}/\chi}|_{r_{AH}}$ is the areal radius evaluated at $r_{AH}$.

{\it Mass of the space-time.} We measure the Misner-Sharp mass  function \cite{MisnerSharp}, which for the metric (\ref{eq:metric}) reads

\begin{eqnarray}
M_{MS} &=& \frac{R}{2}
\left[ 1 + \frac{1}{\alpha^2}(\partial_t R)^{2}  \right.
-2\frac{\beta^r}{\alpha^2}(\partial_t R)(\partial_r R) \nonumber\\
&-& \left. \left( \frac{\chi}{g_{rr}} -\frac{(\beta^r)^2}{\alpha^2} \right)
(\partial_r R)^{2} \right], \label{eq:MisnerSharpMass}
\end{eqnarray}

\noindent where $R=\sqrt{g_{\theta\theta} / \chi}$ is the areal radius. We estimate the ADM mass of spatial slices as the limit $M_{ADM} = \lim_{r \to \infty} M_{MS}$.  The ADM mass measures all the contributions of mass-energy to the space-time, in our case the black hole plus the scalar field, and thus allows one to  separate the contribution of the scalar field and that of the black hole to the total mass-energy of the space-time.

\subsection{Tracking null rays}

Since we want to estimate how the signals sent from near the black hole are received by an observer far away from the black hole, we have to track the trajectory of null rays.

Here we are interested in radial null rays only. We launch a bundle of $m$ null rays whose trajectory in the space-time is given by the points $(r_m,t)$, with initial locations at $r_m(0)=g_m$ near the horizon. Since we consider the rays to be launched radially, the trajectory of these rays is found by equating the radial part of the metric (\ref{eq:metric}) to zero and find that

\begin{equation}
\frac{dr_m}{dt} = -\beta^r(r_m,t) \pm \frac{\alpha(r_m,t)}{\sqrt{\gamma_{rr}(r_m,t)}},
\label{eq:nullrays}
\end{equation}

\noindent which is an equation that determines the radial location of ray $m$ as function of time and metric functions. For simplicity we define $\gamma_{ij}=g_{ij}/\chi$ from (\ref{eq:metric}),  the $\pm$ sign distinguishes between outgoing -the ones we study here- and ingoing rays. The arguments $(r_m,t)$ indicate that metric functions are to be evaluated at the location in space-time of the null geodesic, which in general dos not coincide with points of the numerical domain ${\cal D}_d$ and has to be found by interpolation.  We integrate these equations for $r_m$ simultaneously with the evolution of the space-time itself using the same time integrator as the method of lines.

\section{Kinematics of null rays and tests}
\label{sec:redshift}

We have the trajectory of null-rays, and we now calculate the energy of photons at each point of their trajectories as follows. In a general fashion, for a general 3+1 metric, the motion equation for the location and 4-velocity of test particles is described by the geodesic equations as \cite{Bacchini_2018}

\begin{eqnarray}
\dfrac{dx^i}{dt} &=& \gamma^{ij}\dfrac{u_j}{u^0} - \beta^i,\label{eq: xi}\\
\dfrac{du_i}{dt} &=& -\alpha u^0 \partial_i \alpha + u_k \partial_i \beta^k - \dfrac{u_j u_k}{2u^0}\partial_i \gamma^{jk},
\label{eq: ui}
\end{eqnarray}

\noindent where $u_i = g_{i\mu} u^\mu = g_{i\mu} dx^\mu / d\lambda$ and

\begin{equation}
u^0 = \dfrac{dt}{d\lambda} = \left(\gamma^{jk}u_j u_k + \epsilon\right)^{1/2}/\alpha,
\label{eq: u0}
\end{equation}

\noindent with $\epsilon = 0$ for photons and $\epsilon=1$ for particles with mass and $\lambda$  an affine parameter. The energy measured by a stationary observer with 4-velocity $\left(u^{\mu}\right)_{obs} = \left(|g_{00}|^{-1/2},0,0,0\right)$ along the path of the photon is 

\begin{equation}
E = h\nu = -u_\mu \left(u^\mu\right)_{obs} = -|g_{00}|^{-1/2} u_0 ,
\label{eq: energy}
\end{equation}

\noindent with $u_0 = -\alpha^2 u^0 +\beta^j u_j$ and $g_{00} = -\alpha^2 + \beta_i \beta^i$. Finally the redshift of photons  from the position and time of emission, to position and time of reception is obtained by the expression 

\begin{equation}
1+z = \dfrac{\nu_E}{\nu_R} = \dfrac{E_E}{E_R}
\end{equation}

\noindent  where $E_E$ and $E_R$ are the energy of the photons measured at the position of an emitting source and at the position of a receiving observer respectively.\\

In order to double check our expressions, we reproduce the results for the Schwarzschild space-time using Schwarzschild and isotropic coordinates, which are similar to those used in our evolutions. For a stationary metric  the time coordinate is a cyclic coordinate  for the motion of photons, which implies the zero-component of the 4-momentum of a photon is conserved;  the 1-form corresponding 4-velocity for a far away stationary observer is $(u_\mu)_{obs}=(-\alpha,0,0,0)$ and the energy measured is 

\begin{equation}
E = -u^\mu \left(u_\mu\right)_{obs} = u^0 \alpha.
\label{eq: E stationary}
\end{equation}

\noindent For the Schwarzschild  metric in Schwarzschild coordinates $ds^2 = \left(1-\frac{2M}{\bar{r}}\right)dt^2+d\bar{r}^2/  \left(1-\frac{2M}{\bar{r}}\right) + \bar{r}^2 d\Omega^2$, the system (\ref{eq: xi}, \ref{eq: ui}, \ref{eq: u0}) is reduced for radial trajectories to 

\begin{eqnarray}
\dfrac{d\bar{r}}{dt} &=& 1-\dfrac{2M}{\bar{r}},\label{eq: r Sc}\\
\dfrac{du_{\bar{r}}}{dt} &=& -\dfrac{2M u_{\bar{r}}}{\bar{r}^2},\label{eq: u Sc}\\
u^0 &=& u_{\bar{r}}.\label{eq: Sc}
\end{eqnarray}

\noindent Equation (\ref{eq: u Sc}) can be integrated as function of $\bar{r}$ using the equation (\ref{eq: r Sc}) to obtain 

\begin{equation}
u_{\bar{r}}(\bar{r}) = C\left(1-\dfrac{2M}{\bar{r}}\right)^{-1},
\end{equation}

\noindent where $C$ is a constant of integration. With this, the energy of a photon according to  expression (\ref{eq: energy}) is 

\begin{equation}
E(\bar{r}) = C\left(1-\dfrac{2M}{\bar{r}}\right)^{-1/2},
\label{eq:energy Sc}
\end{equation}

\noindent so the energy measured by the emitting source and the receiving observer located at $ \bar{r}_E $ and $ \bar{r}_R $ respectively is

\begin{equation}
\begin{array}{rcl}
E_E = E(\bar{r}_E) & = & C\left(1-\dfrac{2M}{\bar{r}_E}\right)^{-1/2}, \\
E_R = E(\bar{r}_R) & = & C\left(1-\dfrac{2M}{\bar{r}_R}\right)^{-1/2}, \\
\end{array}
\end{equation}

\noindent and the redshift $z$ is given by

\begin{equation}
1+z = \dfrac{E_E}{E_R} = \sqrt{\dfrac{1-2M/\bar{r}_R}{1-2M/\bar{r}_E}}.
\end{equation}

\noindent Notice that if $\bar{r}_E < \bar{r}_R$  all the signals at the receiver position are redshifted. Now consider the use of isotropic coordinates $ds^2=-\left(\frac{1-\frac{M}{2r}}{1+\frac{M}{2r}}\right)^2dt^2 + \left(1+\frac{M}{2r}\right)^4\left(dr^2+r^2d\Omega^2\right)$, the energy of a photon is

\begin{equation}
E(r) = C\dfrac{1-\frac{M}{2r}}{1+\frac{M}{2r}},
\label{eq:energy iso}
\end{equation}

\noindent which coincides with  Eq. (\ref{eq:energy Sc}) under the coordinate transformation $r=(\bar{r}-M+\sqrt{\bar{r}(\bar{r}-2M)})/2$, which transforms from Schwarzschild to isotropic coordinates. \\

\begin{figure}
\includegraphics[width=8cm]{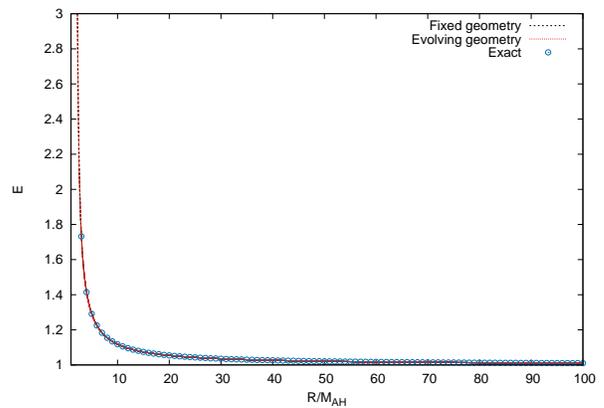} 
\caption{Energy of a photon as function of  areal radius for three cases. When the space-time described in isotropic coordinates remains fixed (black dotted line),  when the metric evolves (red  line), and when using Schwarzschild coordinates (circles). The horizontal axis is in units of areal radius normalized with the black hole mass. This is a first consistency check of results for the space-time that evolves. }
\label{fig:Energy_photon_R}
\end{figure}

It is time to use the general approach in (\ref{eq: xi},\ref{eq: ui},\ref{eq: u0}), that allows one to calculate the energy for a general metric, including the case of an evolving geometry. In order to check whether the calculation of photon energy works during evolution, we measure the energy of photons for the space-time evolved with the GBSSN equations and our gauge conditions, in vacuum, and compare the energy of photons with that in (\ref{eq:energy Sc}) and (\ref{eq:energy iso}). 
In Figure \ref{fig:Energy_photon_R} we show the energy of photons, calculated on the trajectory of their null paths as function of the areal radius $ R_m(t) = \sqrt {g_ {rr}(r_m,t) / \chi(r_m,t)} $. The coincidence in the three cases used indicates that the energy of photons is being calculated correctly when the space-time is evolving.

\begin{figure}
\includegraphics[width=8cm]{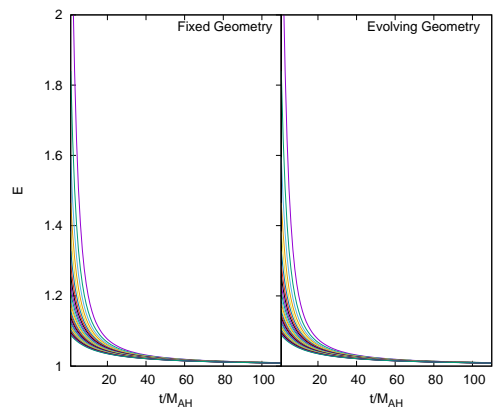} 
\caption{Energy of a bundle of photons as function of the coordinate time $ t $, when spacetime remains fixed in isotropic coordinates (left), when the space-time evolves (right). This is a second consistency check of results for the space-time that evolves.}
\label{fig:Energy_photon_t}
\end{figure}

Similarly, in Figure \ref{fig:Energy_photon_t} we show the energy of a bundle of photons as function of time for the space-time described in isotropic coordinates, and for the space-time evolved with the GBSSN equations in vacuum. These photons are launched from various spatial points $r_m$ with initial radial velocities $ u_{r, m} = (M + 2r_m) / (4r_m^2 (2r_m-M)) $, and their trajectories show gauge independence.

\section{Results for growing/shrinking black holes}
\label{sec:results}

Before exploring the red/blue shift of light emitted from near the black hole, we show the evolution of the black hole while accreting the scalar field. For our analysis, in order to standardize the different scenarios explored, we have fixed a contribution of the scalar field to the total mass of the space-time. When defining the initial profile of the scalar field according to Eq. (\ref{eq:IC_phi}) we choose three values for the width of the wave packet $\sigma=1,5,10$ for both $\kappa=\pm 1$. Another parameter we fix from the general expression (\ref{eq:IC_phi}) for all the cases studied, is the center of the scalar field shell $r_0=40M^{0}_{AH}$, located at a distance of 40 times the apparent horizon mass at initial time. Finally, assuming the ADM mass of the space-time is the same in all cases, we fine-tune the value of $A$ in such a way that the mass of the scalar field is $25\%$ of the mass of the apparent horizon mass at initial time. 

\subsection{Growing and shrinking black holes}
\label{subsec:growshrink}

\begin{figure}
\includegraphics[width=8cm]{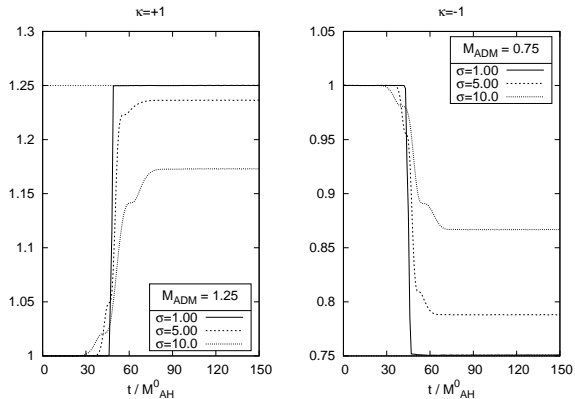}
\caption{Mass of the apparent horizon as function of time for three values of the scalar field initial pulse. In all cases the contribution of the scalar field to the $M_{ADM}$ is 25\%. A small value of $\sigma$ produces a total accretion, whereas larger values allow only a partial accretion of the scalar field. On the left we show the results for the usual scalar field, whereas on the right the results of the phantom case.}
\label{fig:MAH}
\end{figure}

The most significant property of the black hole is the  horizon mass as function of time, which will depend on the amount of scalar field accreted. It is known that the amount of accreted scalar field depends on the size of the wave packet sent toward the black hole, in the fixed geometry case first  \cite{PhysRevD.66.083005}, and confirmed later on for the scalar field when geometry is fully coupled \cite{GuzmanLora2013}. The initial conditions for the scalar field described in Section \ref{subsec:id} include the width of the Gaussian profile, which determines the amount of scalar field accreted and the amount of scalar field escaping from the black hole. In the left panel of Fig. \ref{fig:MAH} we show the mass of the apparent horizon for three values of $\sigma$ for the regular scalar field case, together with the ADM mass of the space-time.   

\begin{figure}
\includegraphics[width=8cm]{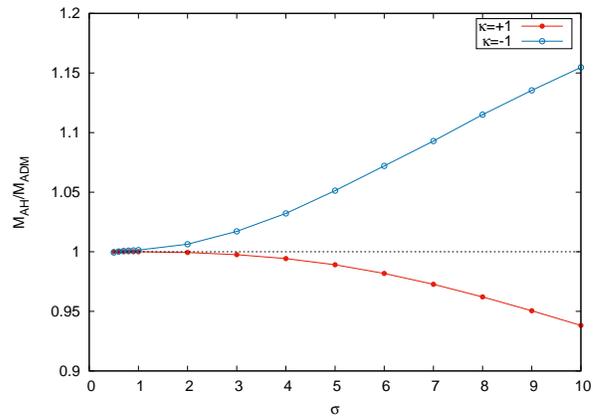}
\caption{ Ratio of absorption of the scalar field as function of $\sigma$. For $\sigma<M^0{}_{AH}$ the scalar field is totally absorbed, whereas for $\sigma >1$ only a portion of the scalar field is accreted. We show the results for the two types of scalar field and $\sigma$ is in units of $M^0{}_{AH}$.}
\label{fig:MvsSigma}
\end{figure}

 A small shell thickness $\sigma=1$ produces the accretion of the total amount of scalar field, which is indicated by the fact that $M_{AH}$ grows from its initial value up to $M_{ADM}$, whereas a wide shell thickness $\sigma=10$ prevents the scalar field from being totally accreted and the horizon mass does not reach $M_{ADM}$, whereas the not accreted scalar field  is scattered and escapes from the black hole. In Figure \ref{fig:MvsSigma} we show the portion of scalar field accreted by the black hole as function of $\sigma$ for the reagular and phantom scalar field cases. We find that for $\sigma < 1$ the absorption is 100\%, whereas for $\sigma > 1$ the absorption decreases as function of $\sigma$. The threshold between full and partial absorption seems to be near $\sigma \sim 1$, consistent with the suggestion that the width of the scalar field pulse plays a similar role as that of the Compton wavelength in the accretion on black holes \cite{PhysRevD.66.083005,GuzmanLora2013}.

\begin{figure}
\includegraphics[width=8cm]{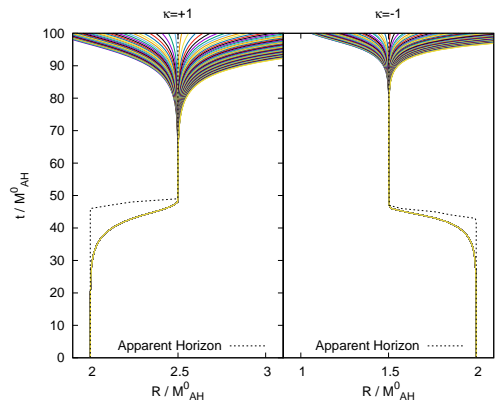}
\caption{Location of the event horizon for the usual scalar field, when the horizon grows, and the phantom scalar field when the horizon shrinks, for the case with $ \sigma = 1 $. The event horizon is the boundary between null rays that escape toward future null infinity and those that evolve toward the black hole singularity. For comparison we also show the apparent horizon radius. }
\label{fig:horizons_with_sigma_eq_1}
\end{figure}

For the sake of illustration and as a consistency check that shows the actual growth and shrink of the event horizon, in Figure \ref{fig:horizons_with_sigma_eq_1} we show the location of the event horizon of the black hole in the case of $\sigma = 1$, when the scalar field is accreted totally by the black hole. The event horizon is located with a fine-tuned set of null geodesics, part of which evolve toward  future null infinity $\scri^{+}$, whereas other part evolve toward the singularity. The event horizon is the surface that separates the two behaviors and corresponds to a single line before $t\sim 70M^{0}_{AH}$. Figure \ref{fig:horizons_with_sigma_eq_1} also illustrates the time scale of the accretion process, which starts when the event horizon starts growing and ends when it stops growing, and if of order of 20 $M^{0}_{AH}$ for $\sigma=1$ and ten times longer for $\sigma=10$.

\subsection{Effects on photon redshift}
\label{subsec:phtonredshift}

We now explore the effects of accretion on the properties of photons. Among the potentially observable properties is the photon's energy while the black hole grows or shrinks. The calculation of this energy uses the general approach in Eqs. (\ref{eq: xi},\ref{eq: ui},\ref{eq: u0}) that we have tested for the vacuum solution, but this time will be used for a black hole that grows or shrinks. 

We define a set of null rays to be tracked as follows. 
We integrate the trajectory of $N$ null rays that are distributed in the domain $D_{geo}=[r_{geo,min},r_{geo,max}]$, with $r_{geo,min}=r_{AH}+0.1$ and $r_{geo,max}=r_{AH}+40.1$ in units of $M^{0}_{AH}$.
The rest mass density of the fluid  $\rho_0$ is defined in $D_{geo}$ and we use such density as the distribution of rays. The initial positions of the $N$ geodesics are $r_{k}$, $k=1,...,N$ where $r_1=r_{geo,min}$ and $r_{k+1}=r_{k}+\Delta r_k$ for $k>1$, and $\Delta r_k=\rho_0(r_1)/\rho_0(r_k)\Delta r_1$ with

\begin{equation}
\Delta r_1 = \frac{1}{\rho_0(r_1) N} \int^{r_{geo,max}}_{r_{geo,min}} \rho_0 dr,
\nonumber
\end{equation}

\noindent provided we calculate the integral numerically. In our analysis we set $N=100$.

\begin{figure}
\includegraphics[width=8cm]{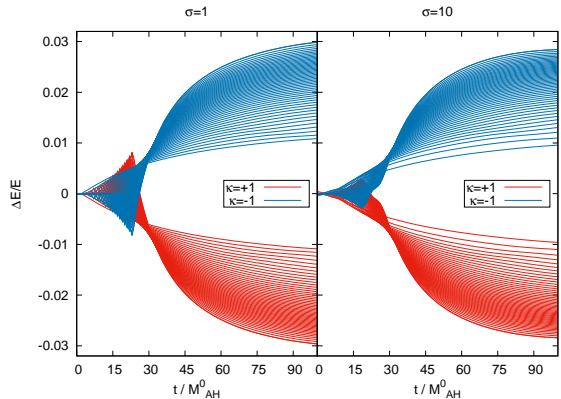}
\caption{Illustration of the relative change in Energy when measured in the presence and absence of scalar field, as a function of time. Red lines correspond to the case of the black hole accreting a usual scalar field that produces a growth of the black hole, and the blue lines to the accretion of the phantom field that shrinks the black hole. These geodesics are launched at coordinate time $t=0$. }
\label{fig. Energy t}
\end{figure}

For a bundle of these photons, the change of photon energy is shown in the Figure \ref{fig. Energy t}, where we show the relative change in energy between the energy measured when the black hole grows or shrinks and the energy measured for zero scalar field, as a function of time. In the case of a growing black hole photons are red-shifted, whereas in the shrinking case photons are blues-shifted. Photons in this Figure are launched at coordinate time $t=0$, by the time the scalar field pulse is at $r_0=40M^{0}_{AH}$.

\subsection{Effects on dust near the black hole}
\label{subsec:effectsondust}

The scalar field distribution distorts the geometry of space-time, so once the appropriate metric functions $\chi$ and $A_{rr}$ are calculated as described in Section \ref{subsec:id}, we assume  the fluid at the initial time is a spherically symmetric, stationary and pressure-less fluid, so that density and velocity distributions are given by the expressions in (\ref{eq:IC_rho}) and (\ref{eq:IC_v}).

\begin{figure*}
\includegraphics[width=17cm]{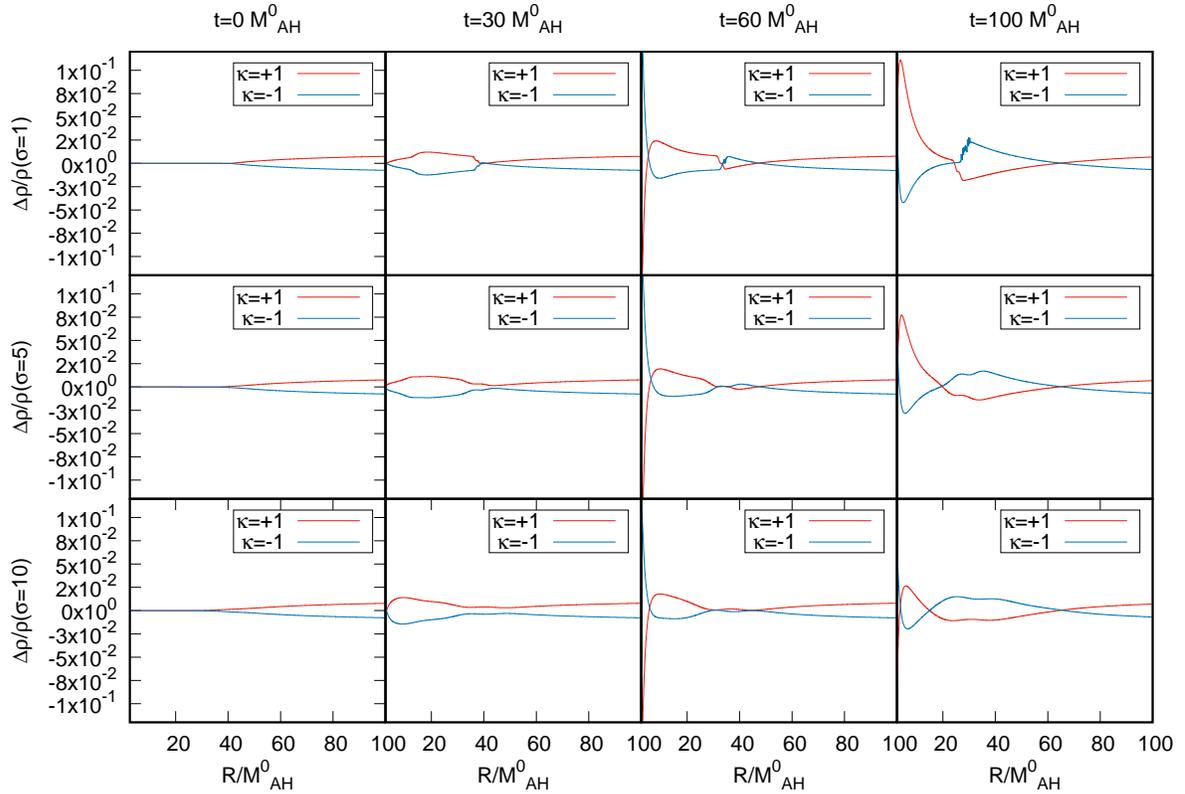}
\caption{Snapshots of the dust density difference in the growth and shrink scenarios, compared with the zero scalar field case. The lines in red correspond to the difference in density due to the accretion of the usual scalar field while the blue to the case of accretion of the phantom scalar field for  three different values of the initial pulse width $\sigma=1,5,10$.}
\label{fig:percent_density}
\end{figure*}

First, we show in the Figure \ref{fig:percent_density} the relative change in the density during the accretion, with respect to the zero scalar field scenario. We use red to indicate the relative difference for the accretion of a regular field, and blue for the accretion of phantom scalar field. As expected, dust has a bigger density near a bigger black hole and smaller for a smaller black hole as illustrated at $t=100M^{0}_{AH}$ when the mass of the final black hole has been stabilized.

\begin{figure}
\includegraphics[width=8cm]{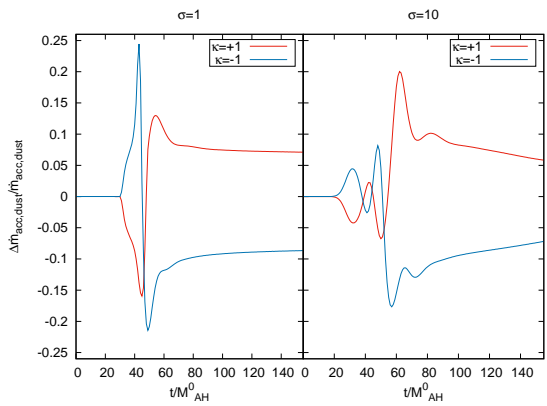}
\caption{Difference in the accretion rate of the dust $\dot{m}_{acc,dust}$ during growth and shrink of the black hole, as function of coordinate time. The lines in red correspond to the difference in dust accretion rate due to a usual scalar field while the blue to a phantom scalar field for two values of the initial pulse width $\sigma=1,10$. This quantity is extracted at the apparent horizon surface.}
\label{fig:percent_macc}
\end{figure}

As expected, the time dependence of the density distribution shown in the snapshots of Figure \ref{fig:percent_density} produces a change in the mass accretion rate with respect to the zero scalar field case. For the sake of illustration, we show the accretion rate of dust during the growth and shrink of the black hole as function of time, measured at the apparent horizon surface in Figure \ref{fig:percent_macc}, for scalar field pulses with $\sigma=1$ and 10.

\subsection{Effects on photon in terms of density}
\label{subsec:phtonredshiftdust}

So far we have described the growth and shrink of the black hole through the accretion of the scalar field. It was also shown how the space-time dynamics, since the dust is a test field, redistributes during the process. Finally, we also described how a bundle of photons distributed according to the dust density profile would redshift when launched at initial coordinate time. In order to estimate the effects of the growth/shrink process on the light emitted from by the dust, and to capture the dynamics of the whole picture, we have to launch a bundle of photons at various time-slices during the growth/shrink  of the black hole.

\begin{figure*}
\centering
\includegraphics[width=8cm]{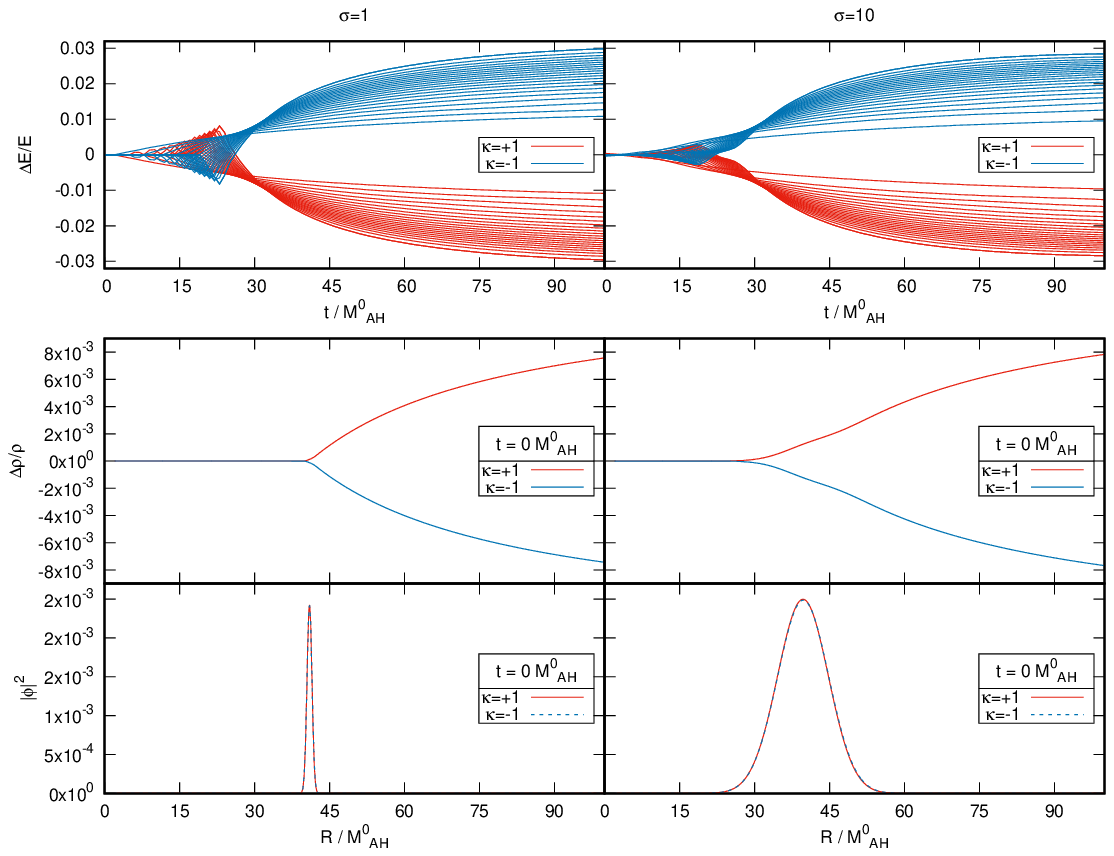}
\includegraphics[width=8cm]{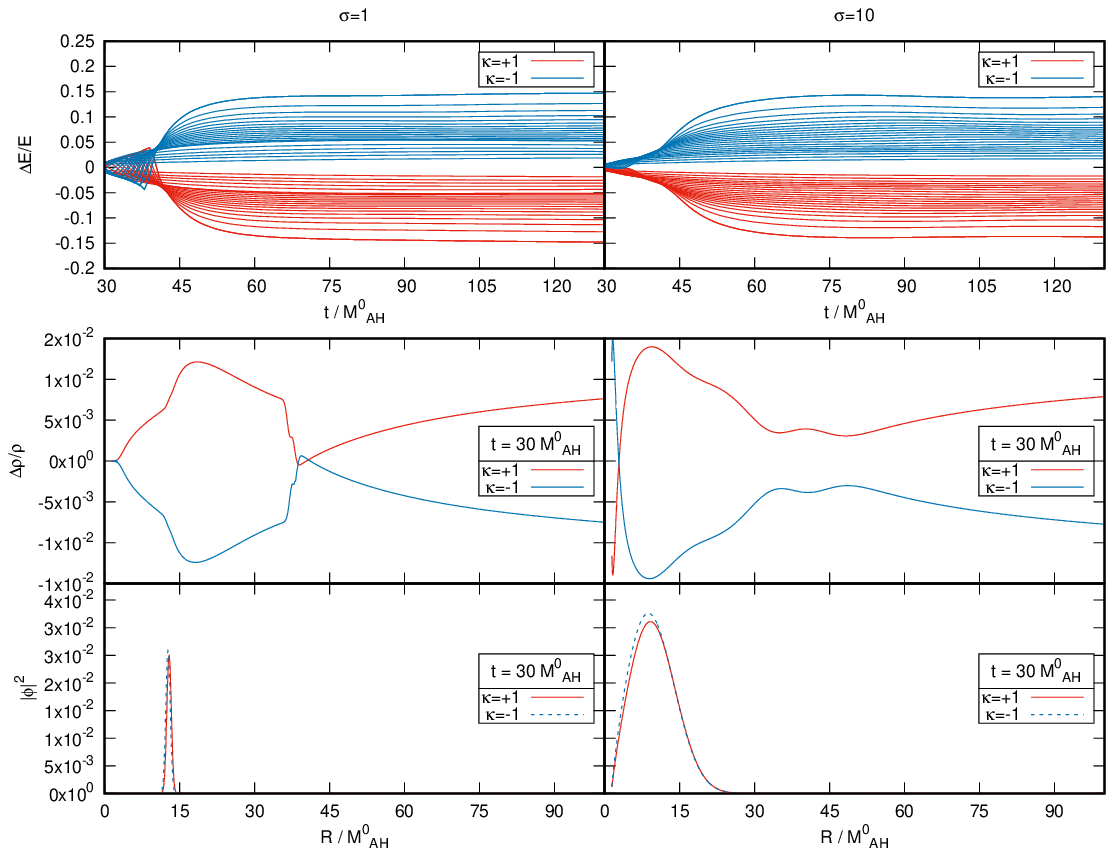}
\includegraphics[width=8cm]{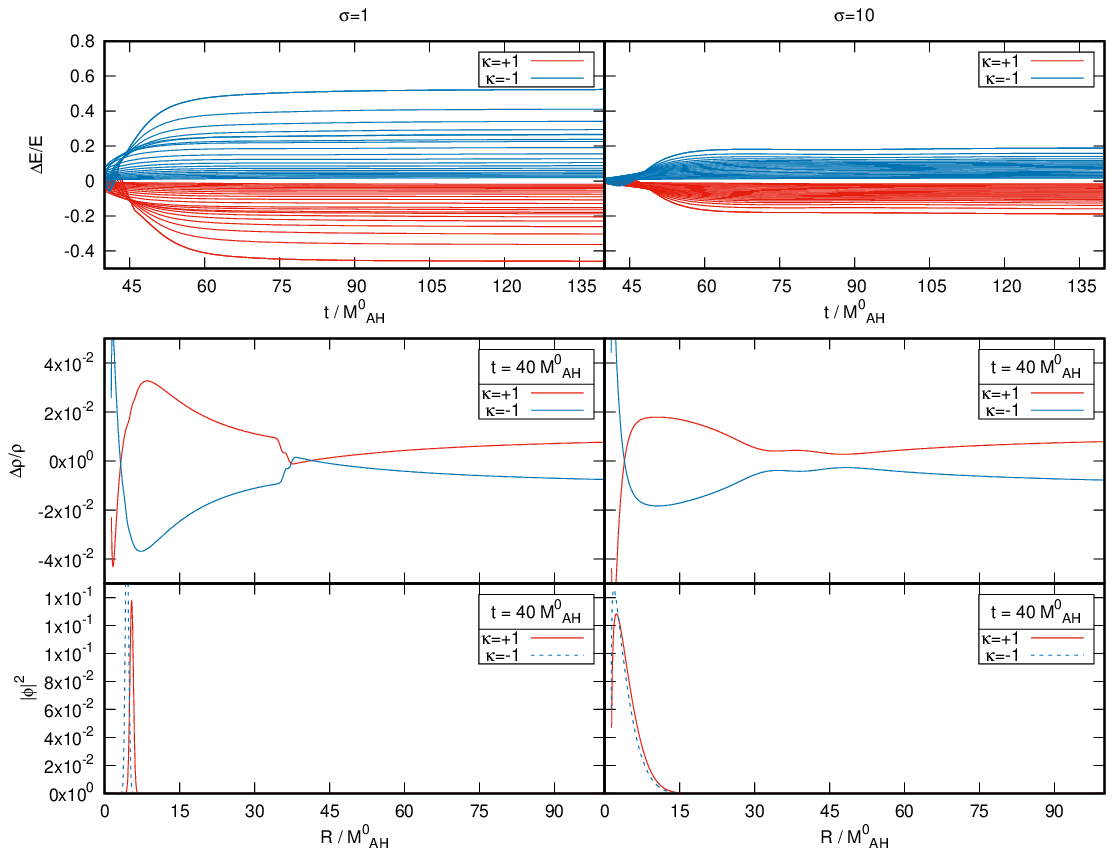}
\includegraphics[width=8cm]{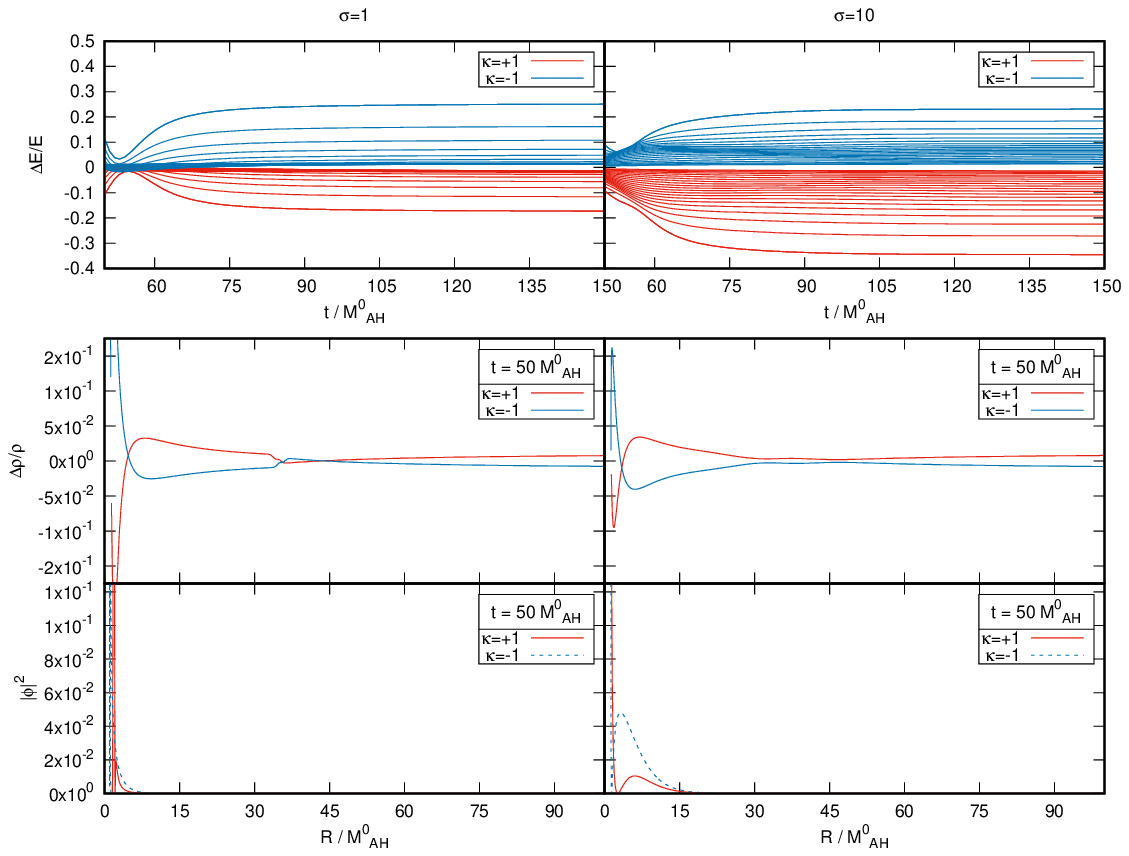}
\includegraphics[width=8cm]{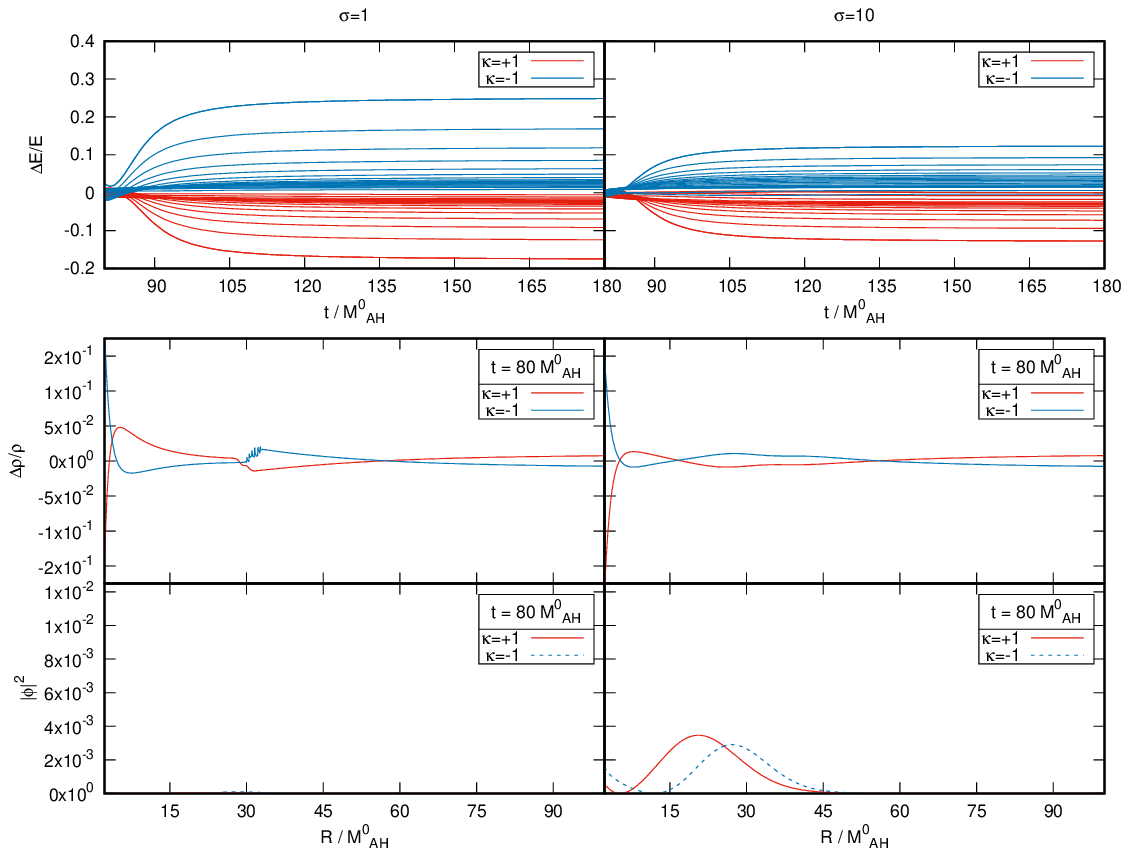}
\includegraphics[width=8cm]{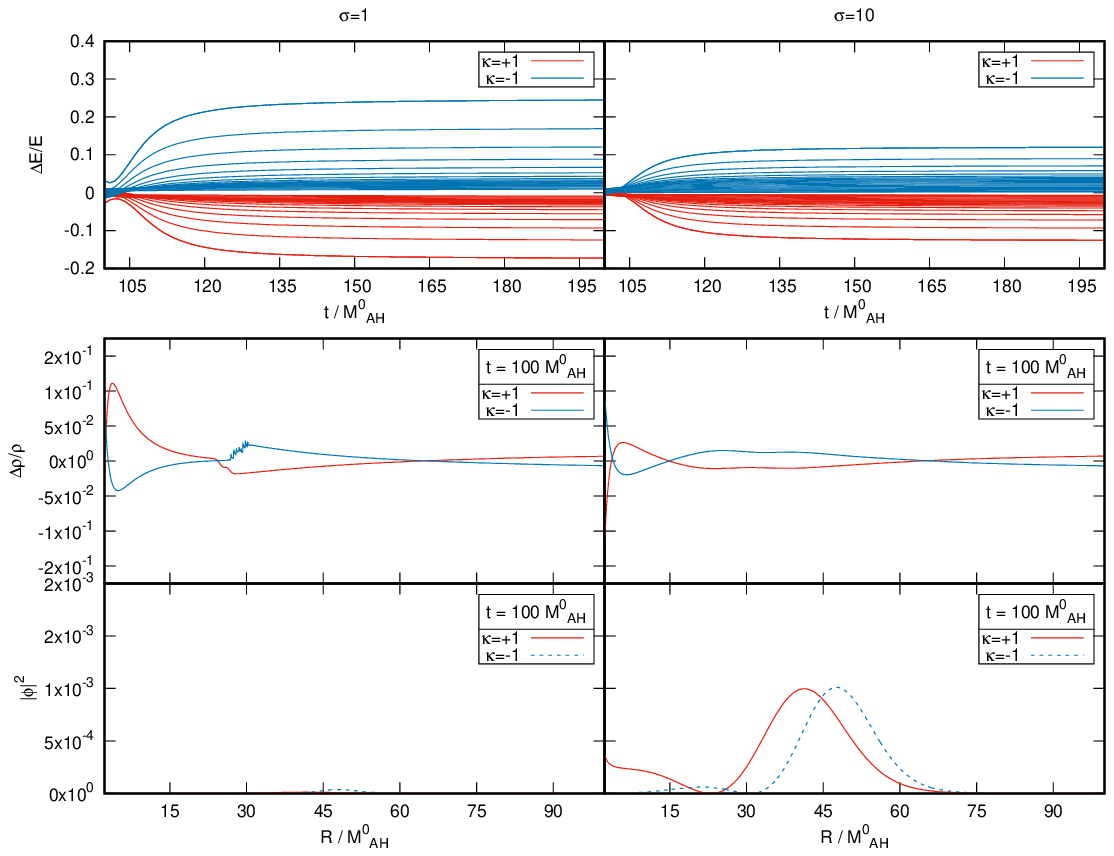}
\caption{In each panel, the upper part shows the relative change in energy for a bundle of photons launched at the indicated time $t$. The graphs in the middle show the relative change in fluid density at launching time $ t$, which illustrate the distortion in the distribution of null rays. In the lower part we show the profile of the scalar field at launching time $ t$, indicating with red the usual scalar field and with blue the case of phantom scalar field, which indicates where the space-time is being more distorted due to the presence of the scalar field at the moment of photon launching. Time is in units of the apparent horizon mass of the black hole at initial time.}
\label{fig:resultsalltimes}
\end{figure*}

We show the results of the process in Fig. \ref{fig:resultsalltimes}. We present six panels, with information about the red/blue shift of photons, the distribution of dust and the distribution of scalar field at the moment when the beam of photons is launched.

In the first row of each panel we show the relative change of photon energy with respect to the case of zero scalar field, for a beam of photons launched at six different coordinate times during the evolution of the space-time, specifically at $t=0,~30,~40,~50,~80,~100M^{0}_{AH}$, for the regular and phantom scalar field cases, and for initial packets with $\sigma=1$ and 10. Notice that the time domain of these plots start at the moment of photon launching, and last for 100 units of time. 

In the middle row of all panels we indicate the relative change of fluid density with respect to the case of zero scalar field at the time when the photons are launched, as function of radius, which illustrates how the distribution of null rays is distorted with respect to the zero scalar field case. Likewise in the third row we plot the square of the scalar field amplitude at the time when the photons are launched, as function of radius at the time when the photons are launched; this illustrates where the scalar field pulse is when the photons are launched and allows one to estimate when they will meet on the way.

Let us comment on the bundle of photons launched at $t=0$. These photons meet the scalar field pulse at about $t\sim 20M^{0}_{AH}$, where $\Delta E/E$ shows a flip of sign and after that the photons travel toward infinity with a nearly constant energy shift. 

When the bundle is launched at $t=30M^{0}_{AH}$ the photons meet the scalar field pulse within the first ten units of time, and $\Delta E/E$ shows a wider fan, with relative energy shifts near $15\%$ with respect to the zero scalar field scenario. 

Photons emitted at $t=40M^{0}_{AH}$ suffer an even bigger relative energy shift between 40 and $60\%$ for a quick accretion with $\sigma=1$ and $20\%$ for a slower accretion with $\sigma=10$. Notice that in the case of $\sigma=10$ the accretion is slower and only $\sim 95\%$ of the scalar field is accreted. 

The bundle launched at $t=50M^{0}_{AH}$, during the time when the tail part of the scalar field is being accreted shows another interesting effect, namely the asymmetry between the accretion of a regular and a phantom field. For $\sigma=1$ the spread of energy is bigger for the shrinking of the black hole, whereas for $\sigma=10$ the growing case shows a bigger spread. The explanation is that the dust responds differently when the black hole is growing than when it is shrinking, that is, the fluid density is different in each case.

Finally we show the results for the cases when the photons are launched at $t=80M^{0}_{AH}$ and $t=100M^{0}_{AH}$, when the black hole mass has stabilized according to Figure \ref{fig:MAH}. The energy shift of photons is nearly the same in the two cases, the distribution of remains time-dependent, due to the post accretion dynamics of the scalar field that keeps distorting the space-time. What is very illustrative from these two cases is that when $\sigma=10$, the scalar field is transmitted through the black hole on a small amplitude pulse that moves outwards and escaped toward infinity.

The potentially observable signature of the process would be trough the variability on the energy shift of photons as function of time, with bundles of photons launched with a more continuos rate than the snapshot in Figure \ref{fig:resultsalltimes}. The movie included in the supplemental material \footnote{http://www.ifm.umich.mx/\~{}guzman/bhblink.html} of the paper illustrates the variability of photon energy shifts during the growing/shrinking process, a dynamical version of Figure \ref{fig:resultsalltimes}.

\section{Conclusions and final comments}
\label{sec:comments}

Based on the solution of Einstein equations sourced by a scalar field with positive and negative energy density, the construction of null-ray trajectories and the energy of photons on those trajectories, we estimated the effects on red/blue shift of photons emitted from regions near the black hole horizon during the accretion of a scalar field.

The accretion of the  scalar field with positive (negative) energy density produces the black hole to grow (shrink). The analysis uses a scalar field characterized by a fixed mass-energy of 25\% of the mass of the initial black hole, distributed in a shell with thickness $\sigma$, implying that the smaller the thickness of the shell the faster the accretion of the scalar field. The time scales of the process we study run from 20 to 100$ M^{0}_{AH}$s and serve to illustrate the effects of horizon growing and shrinking.

Because the space-time is evolving during the accretion process, the red/blue shift of photons during the process depends on the time they are launched at, and thus it is an effect that involves a variability in the frequency shift. In order to capture this effect, a bundle of geodesics is launched at various time slices during the process of accretion. The variability of red/blue shift depends on the thickness of the scalar field pulse, and therefore on how fast or slow the accretion is and how much the black hole grows or shrinks.

The main contribution of our analysis is the method to quantify the energy shift of photons launched from near a black hole while it is growing or shrinking. In the parameter space we studied, the red/blue shift variability, for this scenario of the accretion of matter-energy as big as a quarter of that of the initial black hole mass, can achieve changes up to 60\%.

Some applications of the method may arise. For example, the original astrophysical motivation to study the accretion of scalar fields, include the accretion of ultralight dark matter  onto supermassive black holes (SMBHs) (e.g. \cite{2020PhRvD.101f3532C,PhysRevD.66.083005,PhysRevD.102.063022,chung2021searching,deluca2021tidal,marsh2021astrophysical}), which is ruled by the EKG system of equations. A potential implication of our results is that effects of accretion on SMBHs of this type of matter, could have a fingerprint in frequency shift. 
It can also have applications in the studies of Primordial Black Holes (PBHs) where the accretion during the radiation dominated era is fast and takes place (e. g. \cite{pbh1,pbh2}). 
Another potential application is related to the accretion of exotic cosmic fields that violate the weak energy condition, including phantom fields \cite{2009CQGra..26a5010G,2016PhRvD..93j4044N}.

The general assumptions in our analysis, involving spherical symmetry and the exploration of a reduced parameter space, is the starting point for more realistic configurations on more elaborated and realistic scenarios, like PBHs, or a general scenario of scalar field dark matter accretion onto SMBHs \cite{2020PhRvD.101f3532C}.


\section*{Acknowledgments}
This research is supported by grants CIC-UMSNH 4.9 and 4.23. The runs were carried out in the Big Mamma cluster at the Laboratorio de Inteligencia Artificial y Superc\'omputo, IFM-UMSNH.

\bibliography{bhsaccretion}

\end{document}